\documentclass[12pt]{iopart}

\usepackage{graphicx,amssymb,iopams}
\begin{document}

\title{A GEANT4 Simulation Study of BESIII endcap TOF Upgrade}

\author{Hui Zhang$^1$,Ming Shao$^1$,Cheng Li$^1$,Hongfang Chen$^1$,Yuekun Heng$^2$,Yongjie Sun$^1$,Zebo Tang$^1$,Changsheng Ji$^1$$^,$$^2$,Tianxiang Chen$^1$,Shuai Yang$^1$ and Lailin Xu$^1$}

\address{$^1$ Department of Modern Physics, University of Science and Technology of China. Hefei 230026, China}
\address{$^2$ Institute of High Energy Physics, Chinese Academy of Science, Beijing 100049, China}

\eads{\mailto{swing@ustc.edu.cn (Ming Shao)},\mailto{licheng@ustc.edu.cn (Cheng Li)}}

\begin{abstract}
A GEANT4-based Monte-Carlo model is developed to study the performance of Endcap Time-Of-Flight (ETOF) at BESIII. It's found that the multiple scattering effects, mainly from the materials at the MDC endcap, can cause multi-hit on the ETOF's readout cell and significantly influence the timing property of ETOF. Multi-gap Resistive Plate Chamber (MRPC) with a smaller readout cell structure is more suitable for ETOF detector due to significantly reduced multi-hit rate, from 71.5\% for currently-used scintillator-based ETOF to 21.8\% or 16.7\% for MRPC-based ETOF, depending on the readout pad size used. The timing performance of a MRPC ETOF is also improved. These simulation results suggest and guide an ETOF upgrade effort at BESIII.

\end{abstract}



%\submitto{\MST}
\maketitle

\section{Introduction}
\label{sec:instr}
BESIII \cite{1} is a modern spectrometer located at the upgraded Beijing Electron Positron Collider (BEPCII), which runs in the energy region (2-4.6 GeV) and aims at $\tau$ -charm physics \cite{2}. Particle identification (PID) plays an essential role in the experimental study. One of the main sub-detectors, the time-of-flight (TOF) system, responding for trigger and PID, consists of a barrel and two endcap. Its capability of PID is determined by the flight time difference of particles species and the time resolution of the detector. The current TOF system based on plastic scintillation and photmultiplier is built in 2001 \cite{3}.  The calibration result of endcap TOF system show that the time resolution for electrons in BhaBha events is 148 ps, which is significantly worse than the resolution for mouns in dimu events (110ps, the designed goal). It's also found that scattering in MDC endplate materials can significantly influence the measured times of electrons in BhaBha events \cite{6}.  These findings indicate that the contribution to timing performance from multiple scattering interaction in the endcap material between the main drift chamber (MDC) and the ETOF is important. Improvement is needed for the ETOF system to better meet the BESIII physics goals. The R\&D for this upgrade began in 2010.
Monte-Carlo (MC) simulation serves as an important tool to guide the upgrade, by comparing the performance of TOF based on different technologies and optimizing the prototype design. In this paper a detailed simulation study, including all main features of the BESIII ETOF system, is performed. The GEANT4 \cite{7,8} package (GEANT version 4.09.02.p01), commonly used in high energy experimental physics, is taken as the simulation tool.

\section{Endcap detectors configuration of BESIII}
\label{sec:detecor configuration}
\begin{figure}
\centering
\includegraphics[width=12cm]{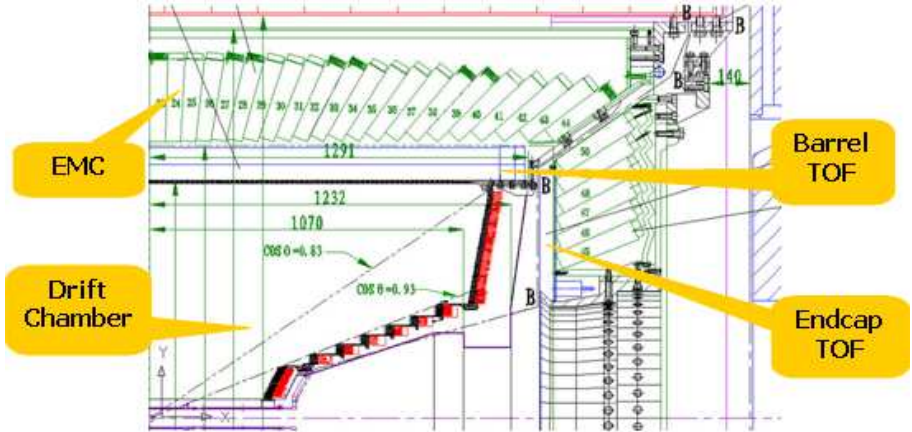}
\caption{Schematic drawing of the TOF in BESIII}
\label{1}
\end{figure}

The BESIII ETOF system is located between a helium-based multilayer MDC and a CsI(Tl) crystal calorimeter (EMC) inside the BESIII spectrometer. The cross-sectional view of the ETOF system is shown in Fig.\ref{1}. Please note although not shown in the figure, there are some materials, such as cables and readout electronics equipment, located between the endplate of the MDC endcap and the ETOF.

\par

\begin{figure}
\centering
\includegraphics[height=8cm]{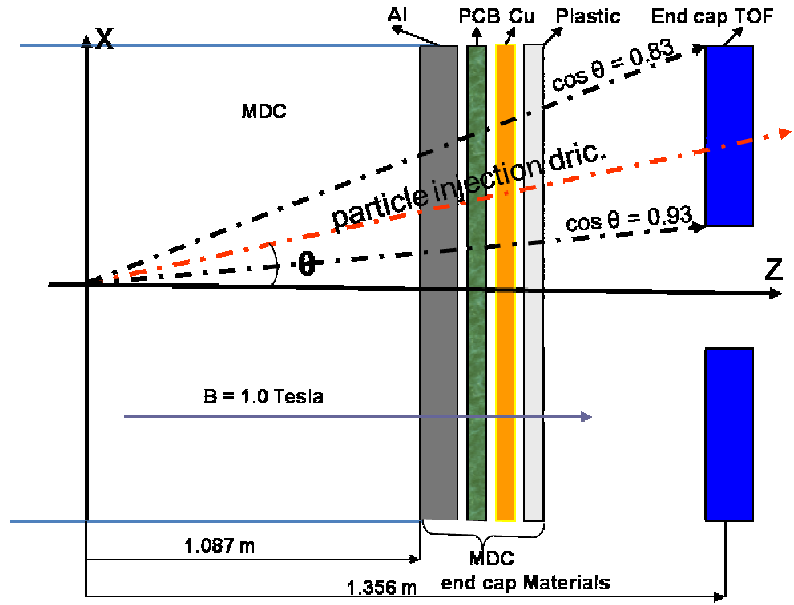}
\caption{Detector layout in simulation}
\label{2}

\end{figure}

Because the description of the detector structure and materials in the full BESIII MC framework is quite complicated and difficult to modify and tune, we develop a simplified model to do this job for the ETOF system. The main structure in our simulation is shown in Fig.\ref{2}. Basically it consists of three parts. The left part is the MDC volume, filled with a gas mixture of 60\% He and 40\% C3H8. The field and sensitive wires in the MDC are not simulated, since their contribution to multiple scattering is small.
\par
The middle part, including the endplate of MDC and the readout electronics and cables, contains the major material budget between the MDC and ETOF. The composition and equivalent thickness of the materials in this part are estimated by using the full BESIII MC framework. Virtual (non-interacting) particles are emitted toward the MDC endcap region with $0.83$ $\leq$ $\cos$$(\theta)$ $\leq$ $0.93$ ( $\theta$ is the polar angle as shown in Fig.\ref{2}). For each step the virtual particle travelled, the material composition and equivalent thickness (along Z direction) are recorded and accumulated until the virtual particle hit the ETOF. We find four major media dominate the material budget, namely aluminum, printed circuit board (PCB), copper and plastic. Their equivalent thicknesses are 21.12mm, 9.77mm, 0.58mm and 9.59mm, respectively.
\par
The right part is the ETOF system, in which the detailed structure depends on the detector technology used and will be addressed in the following part of the paper. The magnetic field of BESIII is a uniform field of 1 Tesla along the Z direction.

\par
\begin{figure}
\centering
\includegraphics[width=12cm]{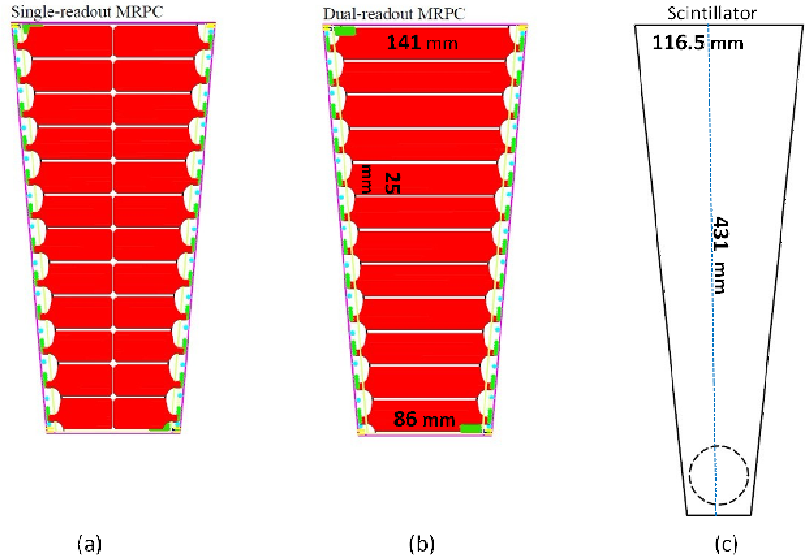}
\caption{Structure of BESIII ETOF module. For (c) the current scintilltor-based ETOF module and (a) the single-readout MRPC-based ETOF module and (b) the dual-readout MRPC-based ETOF module. }
\label{3}
\end{figure}

The current TOF endcap, each consisting of 48 trapezoidal-shaped plastic scintillation (Bicron 404) modules, is located at 1330 mm away from the interaction point (IP) along the beam direction (Z axis in the global BESIII coordinate system, as in Fig.\ref{2}). For each scintillator module, the length is 431mm and the thickness is 48 mm, coving an azimuthal angle range of 7.5${\rm ^o}$ , as shown in Fig.\ref{3}(c). More details can be found in Reference \cite{9}.

\section{Simulation Results of scintillator-based ETOF and Discussion}
\label{results of scintillator}
\par
To simulate the BhaBha events, which play an important role in the offline calibration of BESIII TOF system, electrons with momentum of 1.5GeV/c are emitted from the IP. The injection direction   with respect to the Z axis is chosen to be $\cos$$\theta$ $=0.9$ so that the extrapolated hit point locates near the center of the ETOF module in the radial direction. In the simulation we use standard GEANT4 electromagnetic (EM) physics process including ionization, multiple scattering, bremsstrahlung and the gamma interaction with default settings for production thresholds of secondary particles.
\par
\begin{figure}
\centering
\includegraphics[width=14cm]{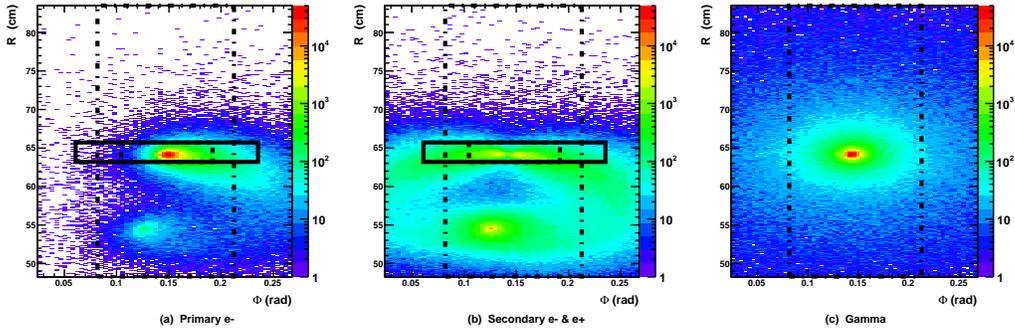}
\caption{Hit position distribution of detected by ETOF. The large rectangle in dashed line shows the boundary of one scintillator-based ETOF module, while the two smaller rectangles in solid/dash line denote the readout-pad boundaries of dual/single readout MRPC module.}
\label{4}
\end{figure}

 Particles hitting the ETOF are electrons, positrons and gamma's. The hit position distributions on ETOF module, of primary electron, secondary electron/position and gamma, are shown in Fig.\ref{4}(a),(b) and (c) respectively. The ETOF module boundaries are also shown in the figures. In order to test the worst case, the extrapolated hit points are chosen to be at the center of an ETOF module (so the hit multiplicity is highest).

%\figurename
\par
\begin{figure}
\centering
\includegraphics[width=8cm]{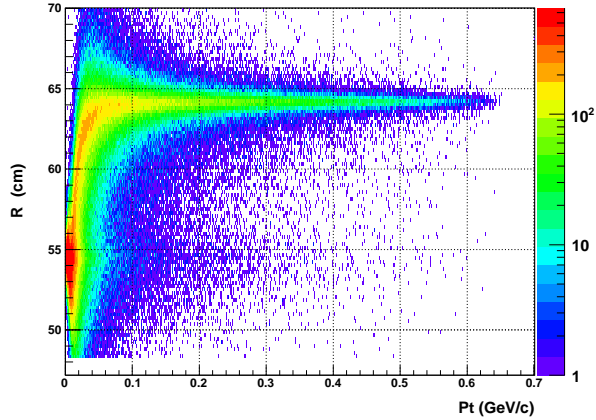}
\caption{The relationship of R with pT of secondary charged particles.}
\label{5}
\end{figure}

To further understand the feature of the secondary charged particles and the effect on ETOF timing, we show in Fig.\ref{5} the radial hit position distribution on ETOF of the secondary electron and positron, as a function of transverse momentum (pT). Two notable bands can be seen in the figure, one with higher pT and R close to 64 cm, while another with very low pT and R close to 54 cm. Combined with Fig.\ref{4}(b), we can conclude a rough scenario that the band near R=64 cm mainly comes from bremsstrahlung gamma conversion to electron/position pair, mostly following the direction of primary electron, while another band located at R=54 cm are those low energy (thus also low \-pT) electrons and positions generated from the EM shower generated by the primary electron or from energetic ionization. In strong magnetic field these low energy electrons or positions basically travel along the field line, thus their hit position on ETOF reflects the position where they are generated, which is R\~54cm from Fig.\ref{2} ($\sim$ 1.1m $\times$sin($\theta$) ).

\par
For each primary electron, a large number of secondary electron, positron and gamma are produced in MDC endcap region causing multi-hit on a scintillator module. More quantitatively, we define the multi-hit rate as the fraction of events with 2 or more hits on one ETOF readout cell. We find the multi-hit rate of scintillator-based ETOF is about 71.5\%. Multi-hit can badly influence the timing performance of ETOF by distorting the output signal shape and amplitude that are hard to be calibrated at offline. Note there is no tracking information available for these secondary particles in the calibration.
\par
\begin{figure}
\centering
\includegraphics[height=8cm]{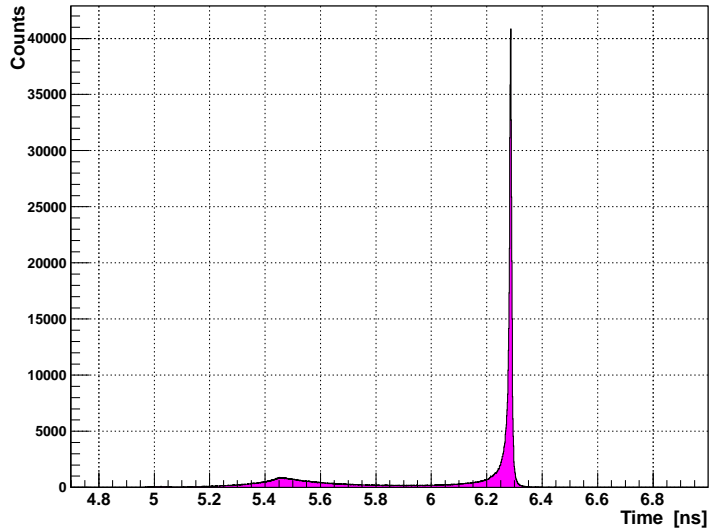}
\caption{The timing distribution of scintillator-based ETOF.}
\label{6}
\end{figure}

To study the effect of the complex hit position distribution structure and the high multi-hit rate on ETOF timing, we make a semi-quantitative calculation by evaluating the hit time and position information.
For each track, the measured arrival time is the sum of particle flight time from the IP all the way to ETOF and the signal's transmission time to the readout end. We assume a simple one-dimensional linear timing dependence on the hit position. In the scintillator, the signal transmission time can be expressed as (R-R0)/v, where R0 is the radial position of the photomultiplier, and v is the effective transmission velocity in scintillator. R0 is 47.2 cm in current BESIII ETOF design and v-1 is measured to be 80 ps/cm \cite{6}. In each event, the earliest arrival time at the photomultiplier among all hits is taken as the measured TOF. The TOF distribution is shown in Fig.\ref{6}. Beside the nominal TOF peak at around 6.2-6.4 ns caused by primary electrons, another peak with much smaller TOF around 5.5 ns is clearly visible in Fig.\ref{6}. The left peak is about 0.9 ns earlier in time than the nominal peak - this is understood as a consequence of secondary electron/positron band at low energy and low R in Fig.\ref{5}. In Fig.\ref{5}, the difference in R for the two main bands is ~10 cm. Consider the transmission velocity of fluorescent photons in scintillator, their arrival time difference will be 0.8 ns, agreeing well with the Fig.\ref{6}. The difference in flight time from the IP to ETOF is relatively small, since electron or position travels at a velocity very near to the speed of light. Our simulation also confirms this point. Furthermore, similar feature was observed with the full BESIII MC framework (including fluorescent photon generation and transmission), as well as the calibration results with experimental BhaBha data collected by BESIII \cite{6}. The consistency of these results validates the reliability of our simulation.

\par
The simulation results of scintillator-based ETOF show that the multiple scattering effects, mainly from the MDC endcap materials, can cause multi-hit on the ETOF's readout cell and make the hit position distribution structure on the ETOF's readout cell complex by producing secondary particles in the materials and significantly influence the performance of ETOF. Furthermore, the complex hit position distributions also indicate that the tracking accuracy is also worse in the endcap region than that in the barrel region, which consequently makes the position-dependent time calibration difficult.
\par
To reduce the multi-hit probability and simplify the hit position distribution structure, smaller readout cell size is favored. However, reducing module size means increasing readout channels, which is not suitable for scintillator-based ETOF since the PM dedicated in strong magnetic field is rather expensive. The MRPC, first developed by LHC-ALICE TOF collaboration \cite{10}, is a new type gaseous detector with good time resolution, high detection efficiency. Furthermore, it can be designed with a smaller readout cell structure and the cost of each readout cell is much lower. It is considered a suitable candidate for the upgrade of BESIII ETOF system.
\par
\begin{figure}
\centering
\includegraphics[width=12cm]{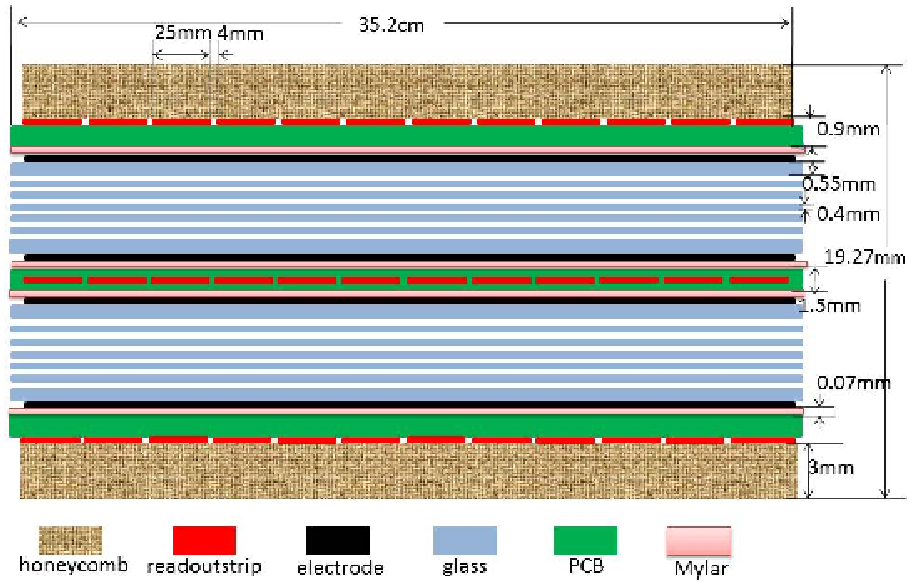}
\caption{Side view of an MRPC module for BESIII ETOF upgrade. }
\label{7}
\end{figure}

The R\&D for such an upgrade began in 2010. In each endcap there are 36 trapezium-shaped MRPC modules. Fig.\ref{7} shows the top and side view of an MRPC module appropriate for BESIII. It has a double-stack structure with twelve gaps. Floating glass sheets are used as the resistive plates. The thicknesses are 0.4 mm and 0.55 mm for the inner and outer glass, respectively. The gap between the glasses is 0.22 mm. The thickness of the honeycombs, readout pads, PCB and Mylar boards are also shown in Fig.\ref{7}. The MRPC is placed in a aluminum box with a thickness of 1 mm which is filled with a standard gas mixture for MRPC. The component of the gas is 90\% Freon + 5\% SF6 + 5\% C4H10 \cite{11}.

Two types of MRPC modules with different readout pad structure are designed for the upgrade. The dual-readout MRPC module has 12 readout pads of 2.5 mm wide, with the length ranging from 8.6 cm to 14.1 cm. Signals are readout at both ends of the pad. The single readout MRPC module has a similar structure except that each readout pad is divided into two from the center (see Fig.\ref{3}). In order to study the improvement with this upgrade, the scintillator-based ETOF system is replaced with the proposed MRPC based ETOF system in the simulation setup shown at Fig.\ref{2}.

\section{Simulation Results of MRPC based ETOF and Comparison}
\label{result of MRPC}
\par
 The hit position distributions on MRPC based ETOF module, of primary electron, secondary electron/positron are shown in Fig.\ref{4}(a) and (b) respectively. The module boundaries are also shown in the figures like that of scintillator-based ETOF case. Note in Fig.\ref{4}(c) the MRPC pad boundaries are not shown since MRPC are intrinsic not sensitive to gamma (usual efficiency $<$ 0.1\%).
 \par
 It's found that the multi-hit rate is 21.8\% or 16.7\% for MRPC ETOF with dual-readout or single-readout module design, dropping significantly from scintillator-based ETOF's 71.5\%. Note the readout cell of MRPC based ETOF is much smaller than that of scintillator-based ETOF and all three components in Fig.\ref{4} (primary electron, secondary electron/position and gamma) contribute for a scintillator while gamma hits are not accounted for MRPC.
 \par
With a significantly reduced multi-hit event rate and simpler hit position distribution structure, one can expect better timing performance and easier calibration for a MRPC-based ETOF rather than a scintillator-based ETOF. To compare the timing performance of both types of ETOF, we also make a semi-quantitative calculation for MRPC based ETOF in a similar way as the scintillator.
\par
\begin{figure}
\centering
\includegraphics[width=14cm]{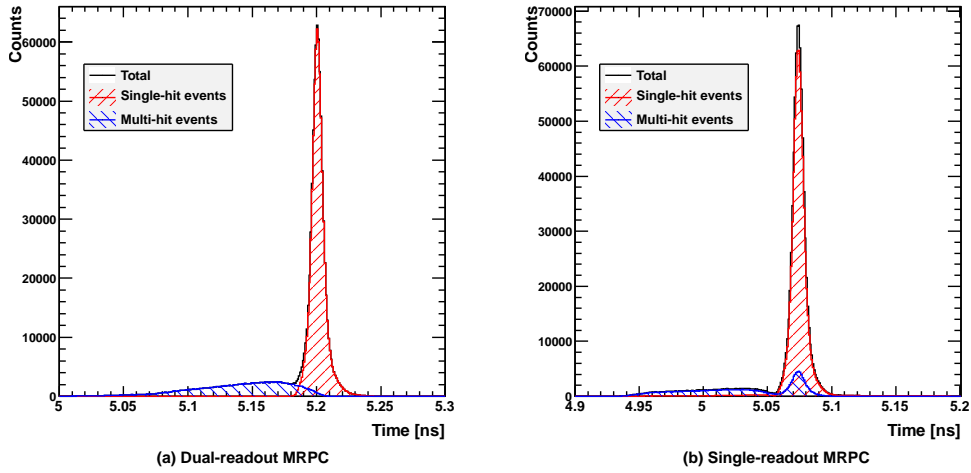}
\caption{The timing distribution of MRPC-based ETOF, for (left) the dual-readout MRPC design and (right) the single-readout MRPC design.}
\label{8}
\end{figure}

 On the MRPC readout pad, electric pulses propagate to the FEE-fed end. For each track, the signal transmission time on the readout pad is calculated as L/v, where L is the distance between the hit and the feed-out end, and v is the propagation velocity of the electric pulse. For MRPC modules used in BESIII ETOF, v-1 is measured to be 45 ps/cm \cite{12,13}, near twice faster than that in an ETOF scintillator. Also in each event, only the earliest signal arrival time is taken as the measured TOF. For single-readout MRPC the signal is chosen to be feed-out from the left end, while for dual-readout MRPC, the time measured is chosen as the average measurement from both sides. The TOF distribution of a MRPC ETOF is shown in Fig.\ref{8}. It's obvious that the false timing from secondary particles is greatly suppressed compared to the scintillator ETOF case.
 \par
\begin{figure}
\centering
\includegraphics[height=9cm]{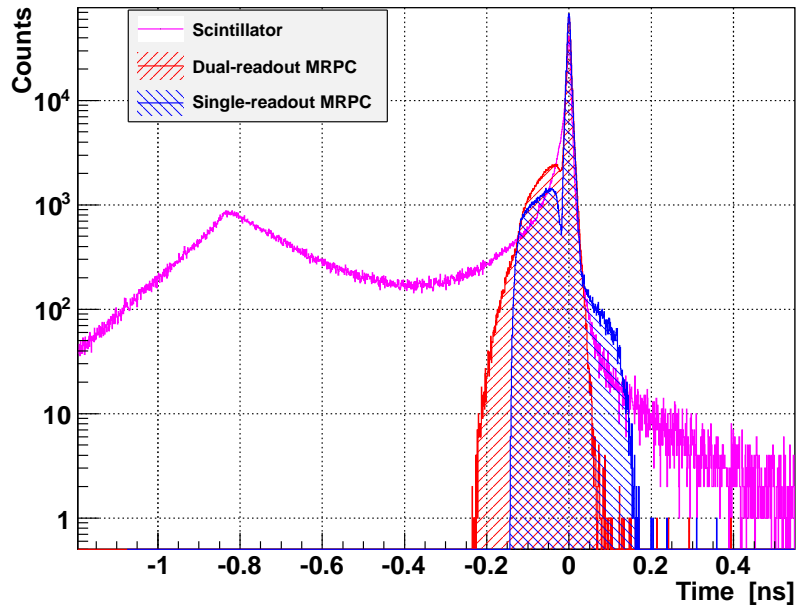}
\caption{Comparison of timing of scintillator-based and MRPC-based ETOF. (Peaks are moved to zero ns)}
\label{9}
\end{figure}

We make a comparison of timing performance from all three kinds of ETOF design (two detector technology, and two readout pad design for MRPC) and show the result in Fig.\ref{9}. Again we see clear improvement with MRPC-based ETOF. For the two kinds of MRPC readout design, the single-readout method has better timing property due to less multi-hit rate compared to the dual-readout one. However, one should know for single-readout MRPC, the timing is hit position dependent thus requires precise tracking which is not easy at BESIII endcap region. Dual-readout MRPC basically does not need very good tracking since the hit position uncertainty cancels out in time averaging from both ends. Further investigation need more detailed simulation and experimental efforts and is beyond the scope of this paper.

\par
To further quantify the difference in timing for the three kinds of ETOF design, we calculated the fraction of events in which the time difference (with respect to Time=0 in Figure \ref{9}) larger than 30 ps, 50 ps, 80 ps, 100 ps and 150 ps. The results are shown in Table.\ref{tabone}. We can clear see that the upgrade can further reduce on the adverse influence of the multi-hit on the timing of ETOF.
\par

\begin{center}
\begin{table}
\centering
\caption{\label{tabone}}

%\begin{indented}
\lineup
\item[]\begin{tabular}{@{}*{6}{l}}

\br
Types&$\Delta$t$\geq$30ps&$\Delta$t$\geq$50ps&$\Delta$t$\geq$80ps&$\Delta$t$\geq$100ps&$\Delta$t$\geq$150ps\cr
%Types & $\Delta$t$\geqq$ 30ps  & $\Delta$t$\geqq$ 50ps  & $\Delta$t$\geqq$ 80ps  & $\Delta$t$\geqq$ 100ps  & $\Delta$t$\geqq$ 150ps\
\mr
%\hline
Scintillator & 40.0\% & 36.2\% & 33.3 \% & 32.1 \% & 30.0\% \\
Dual-readout MRPC & 16.3\% & 11.6\% & 5.9\% & 3.3\% & 0.5\% \\
Single-readout MRPC&9.4\% & 7.1\% & 3.7\% & 1.9\% & 0.0\% \\

\br
\end{tabular}%
%\end{indented}
\end{table}
\end{center}

\section{Conclusion}
\label{conclusion}

We have developed a GEANT4-based MC model to study the timing property of the two types of ETOF design - based on scintillator or MRPC. The simulation results show that multiple scattering process in the material budget between MDC endcap and ETOF will produce a large number of secondary particles that will affect the timing of the plastic scintillator used in current BESIII ETOF system, causing a high multi-hit rate (71.5\%) and a double-peak structure in the TOF spectrum for BhaBha events. The timing peak with TOF abnormally small is contributed by the secondary particles with very low energy and is about 0.9 ns earlier to the nominal timing peak. These results are consistent with the offline calibration results with experimental BhaBha data. The multi-hit rate of MRPC is much lower (21.8\% for Dual-readout MRPC and 16.7\% for single-readout MRPC design), and the timing performance of MRPC is also better.

\section*{Acknowledgment}

This work was supported by National Science Foundation of China, no.10970003.


\section*{References}

\begin{thebibliography}{00}

\bibitem{1}{M. Ablikim {\it et al.},Design and Construction of the BESIII Detector,{\it Nucl. Instr. and Meth.} {\bf A 614}(2010) 345.}
\bibitem{2}{D. Asner {\it et al.},Physics at BES-III,{\it Int. J. Mod. Phys.} {\bf A 24}(2009)Suppl. 1.}
\bibitem{3}{X .Li, {\it et al.}, Monte Carlo Simulation of BES III End-Cap TOF,{\it High Energy Physics . Nuclear Physics.} {\bf 29}(2005)586 (in Chinese).}
\bibitem{4}{Z.B. Tang, {\it et al.},R\&D of the Endcap TOF Detector for BESIII,{\it High Energy Physics. Nuclear Physics.}{\bf 30}(2006) 445.(in Chinese).}
\bibitem{5}{S.H. An, {\it et al.}, Beam Test of the Time Resolution of the BESIII Endcap TOF,{\it High Energy Physics. Nuclear Physics.}{\bf 29}(2005) 775(in chinese).}
\bibitem{6}{C. Zhao {\it et al.}, Time calibration for the end cap TOF system of BESIII, {\it Chinese Physics} {\bf C 35(1)} (2011) 72.}
\bibitem{7}{S. Agostinelli, {\it et al.}, GEANT4-a simulation toolkit, {\it Nuclear Instrument and Method.}{\bf A 506}(2003) 250. }
\bibitem{8}{J. Allison, {\it et al.},Geant4 developments and applications,{\it  IEEE Trans. Nuclear Science} {\bf NS-53}(2006) 270. }
\bibitem{9}{Y. Liu {\it et al.},A GEANT4-based simulation model for the BESIII endcap time-of-flight system,{\it Nuclear Instrument and Method}{\bf A 629} (2011) 87. }
\bibitem{10}{ALICE Time-of-Flight Proposal.. http://alice.web.cern.ch/Alice/TDR/alice-tof.ps. }
\bibitem{11}{Y. Sun, {\it et al.},A prototype MRPC beam test for the BESIII ETOF upgrade {\it Chinese Physics}{\bf C 35(5)}(2012) 429. }
\bibitem{12}{A. Akindinov {\it et al.},Results from a large sample of MRPC-strip prototypes for the ALICE TOF detector,{\it Nuclear Instrument and Methods} {\bf A 532} (2004) 611. }
\bibitem{13}{J. WU {\it et al.}, The performance of the TOFr tray in STAR,{\it Nuclear Instrument and Methods}{\bf A492} (2005) 344 }
\end{thebibliography}
\end{document}